\documentclass[twocolumn,showpacs,prd]{revtex4}
\usepackage{mathrsfs}
\usepackage{longtable,lscape}
\usepackage{txfonts}
\usepackage{amssymb}
\usepackage{indentfirst}
\usepackage{graphicx,,booktabs}
\usepackage{color}
\usepackage{amssymb}

\begin{document}

\title{Two charged strangeonium-like structures observable in the $Y(2175) \to \phi(1020)\pi^{+} \pi^{-}$ process}
\author{Dian-Yong Chen$^{1,3}$}\email{chendy@impcas.ac.cn}
\author{Xiang Liu$^{1,2}$\footnote{Corresponding author}}\email{xiangliu@lzu.edu.cn}
\author{Takayuki Matsuki$^4$}\email{matsuki@tokyo-kasei.ac.jp}

\affiliation{
$^1$Research Center for Hadron and CSR Physics,
Lanzhou University and Institute of Modern Physics of CAS, Lanzhou 730000, China\\
$^2$School of Physical Science and Technology, Lanzhou University, Lanzhou 730000,  China\\
$^3$Nuclear Theory Group, Institute of Modern Physics of CAS, Lanzhou 730000, China\\
$^4$Tokyo Kasei University, 1-18-1 Kaga, Itabashi, Tokyo 173-8602, Japan}
\date{\today}

\begin{abstract}
Via the Initial Single Pion Emission (ISPE) mechanism, we study the $\phi(1020)\pi^{+}$ invariant mass spectrum distribution of $Y(2175) \to \phi(1020)\pi^{+} \pi^{-}$. Our calculation indicates there exist a sharp peak structure $\left(Z_{s1}^+\right)$ close to the $K\bar{K}^\ast$ threshold and a broad structure $\left(Z_{s2}^+\right)$ near the $K^\ast\bar{K}^\ast$ threshold. In addition, we also investigate the $\phi(1680) \to \phi(1020)\pi^{+} \pi^{-}$
process due to the ISPE mechanism, where a sharp peak around the $K\bar{K}^\ast$ threshold appears in the $\phi(1020)\pi^{+}$ invariant mass spectrum distribution. We suggest to carry out the search for these charged strangeonium-like structures in future experiments, especially Belle II, Super-B and BESIII.

\end{abstract}

 \maketitle

\section{introduction}\label{sec1}

As one of the newly observed $XYZ$ states, $Y(2175)$ was first reported by
the BaBar Collaboration, where an enhancement structure with mass
$M=2175\pm10\pm15$ MeV and width $\Gamma=58\pm16\pm20$ MeV was observed in
the $\phi f_0(980)$ invariant mass spectrum of $e^+ e^-\to \phi f_0(980)$
via the initial state radiation (ISR) mechanism \cite{Aubert:2006bu}.
Furthermore, $Y(2175)$ was confirmed by BES-II in $J/\psi\to \eta \phi
f_0(980)$ \cite{:2007yt} and by Belle in the $e^+ e^- \to \phi\pi^+\pi^-$ and
$e^+ e^-\to \phi f_0(980)$ processes \cite{Shen:2009zze}. Although
experimental observation of $Y(2175)$ is mainly due to the analysis of the
$\phi\pi^+\pi^-$ and $\phi f_0(980)$ invariant mass spectra,
experimentalists have also performed the search for $Y(2175)$ by the other decay
channels. The BES-II Collaboration has indicated that no evidence of $Y(2175)$
is seen by analyzing the $K^{*0}\bar{K}^{*0}$ invariant mass spectrum in
$J/\psi\to \eta K^{*0}\bar{K}^{*0}$ \cite{:2009gza}. Later, BaBar has observed
an enhancement structure around 2127 MeV in the $\phi\eta$ invariant mass
spectrum of $e^+e^-\to \phi\eta$ via the ISR mechanism
\cite{Aubert:2007ym}.

The observation of $Y(2175)$ have stimulated theorists' interest in
revealing its underlying structures. By a relativized quark model with
chromodynamics \cite{Godfrey:1985xj}, Godfrey and Isgur predicted the
masses of $2^3 D_1$ and $3^3 S_1$ states close to the mass of $Y(2175)$,
which seems to support $Y(2175)$ as a vector strangeonium. However, vector
strangeonium assignment with $3^3 S_1$ can be fully excluded since the
calculated total width of this state is about 380 MeV
\cite{Barnes:1996ff,Barnes:2002mu}, which is far larger than the width of
$Y(2175)$. Right after the observation of $Y(2175)$, Ding and Yan studied the
decay behavior of $Y(2175)$ assuming it as a $2^3 D_1$ $s\bar{s}$ state
\cite{Ding:2007pc} and calculated the total width of $Y(2175)$ to be 167.21
MeV by the $^3P_0$ model or 211.9 MeV by the flux tube model, which are
larger than the width of $Y(2175)$ to some extent. Under the $2^3 D_1$
$s\bar{s}$ strangeonium scenario, the main decay modes of $Y(2175)$ are
$K\bar{K},\, K^*K^*,\,K(1460)K,\,h_1(1380)\eta$ \citep{Ding:2007pc}. The
authors of Ref. \cite{Coito:2009na} applied the Resonance Spectrum
Expansion (RSE) model to study excited $1^{--}$ $s\bar{s}$ states, where
in the $q\bar{q}$ sector both the $^{3}S_1$ and $^{3}D_1$ states are
coupled. A dynamical resonance pole at $(2186-246i)$ MeV was found in Ref.
\cite{Coito:2009na}, which obviously shows that such a resonance with huge
width is inconsistent with the experimental data of $Y(2175)$. As
presented in Ref. \citep{Coito:2009na}, further improvements of the model
are needed to the calculation of the pole mass.
Other than conventional $s\bar{s}$ assignment, there exist a couple of exotic
explanations to $Y(2175)$, which include $s\bar{s}g$ hybrid state
\cite{Ding:2006ya}, $K\bar{K}\phi$ three-body system
\cite{MartinezTorres:2008gy}, tetraquark state
\cite{Wang:2006ri,Chen:2008ej} and $\Lambda\bar{\Lambda}$ molecular state
\cite{Klempt:2007cp,Qiao:2005av}.

The observation of $Y(2175)$ is tempting us to relate it to the
observed $Y(4260)$ and $\Upsilon(10860)$ due to some common peculiarities
existing in the experiments of $Y(2175)$, $Y(4260)$ and $\Upsilon(10860)$.
Before observing $Y(2175)$, the BaBar Collaboration once reported an
enhancement named as $Y(4260)$ in the $J/\psi\pi^+\pi^-$ invariant mass
spectrum of $e^+e^-\to \gamma_{ISR} J/\psi\pi^+\pi^-$
\cite{Aubert:2005rm}. In 2007, the Belle Collaboration found anomalous
partial width of $\Upsilon(10860)\to \Upsilon(1S,2S)\pi^+\pi^-$, which is
$2\sim 3$ orders larger than those of $\Upsilon(nS)\to
\Upsilon(mS)\pi^+\pi^-$ ($n=2,3,4$ and $m<n$) \cite{Abe:2007tk}.

Comparing the experimental phenomena of $Y(2175)$, $Y(4260)$ and
$\Upsilon(10860)$, we notice the similarities among these particles.
Firstly, $Y(2175)$, $Y(4260)$ and $\Upsilon(10860)$
are produced from the $e^+e^-$ collision, which indicates that their quantum
numbers are $J^{PC}=1^{--}$. Secondly, the dipion transitions of
$Y(2175)$, $Y(4260)$ and $\Upsilon(10860)$ were observed. Thirdly, there
exist some anomalous phenomena in the $e^+e^-$ collisions at several
typical energies $\sqrt{s}=2175$ MeV, $4260$ MeV and $10860$ MeV, which
correspond to the relevant observations of $Y(2175)$, $Y(4260)$ and
$\Upsilon(10860)$. In Eq. (\ref{common}), we give a brief summary
of our observation, i.e.,
\begin{eqnarray}
e^+e^-\Rightarrow
\left\{
  \begin{array}{cccc}
    Y(2175) & \rightarrow  & \phi(1020)\pi^+\pi^-&\mathrm{strange}\\
    Y(4260) & \rightarrow  & J/\psi\pi^+\pi^-&\mathrm{charm}\\
    \Upsilon(10860)& \rightarrow  & \Upsilon(1S,2S)\pi^+\pi^-&\mathrm{bottom}\\
  \end{array}
\right. ,\label{common}
\end{eqnarray}
which seems to show a complete series of flavors. Thus, the ideas that arise when
studying $Y(2175)$, $Y(4260)$ and $\Upsilon(10860)$ can be shared with each other,
which provides new insight into these peaks and further reveals the properties of
$Y(2175)$, $Y(4260)$ and $\Upsilon(10860)$.

Recently, the Belle Collaboration announced two charged bottomonium-like
structures $Z_b(10610)$ and $Z_b(10650)$ observed in the
$\Upsilon(1S,2S,3S)\pi^\pm$ and $h_b(1P,2P)\pi^\pm$ invariant mass spectra
of $\Upsilon(10860)$$\to$$\Upsilon(1S,2S,3S)\pi^+\pi^-$,
$h_b(1P,2P)\pi^+\pi^-$ \cite{:2011pd}. In Ref. \cite{Chen:2011pv}, the Initial Single
Pion Emission (ISPE) mechanism, a unique mechanism, was introduced in the
$\Upsilon(10860)$ hidden-bottom dipion decay. The ISPE mechanism can
naturally explain why the $Z_b(10610)$ and $Z_b(10650)$ enhancements
close to the $B\bar{B}^*$ and $B^*\bar{B}^*$ thresholds, respectively,
appear in the $\Upsilon(1S,2S,3S)\pi^\pm$ and $h_b(1P,2P)\pi^\pm$
invariant mass spectra. Emphasized in Ref. \cite{Chen:2011pv}, extending
the ISPE mechanism to study the hidden-charm dipion decays of higher
charmonia is also an interesting research topic. The numerical result
indicates that there exist peak structures near the $D\bar{D}^*$ and
$D^*\bar{D}^*$ thresholds, which can be accessible in future experiments
\cite{Chen:2011xk}. In addition, via the ISPE mechanism,
the charged bottomonium-like structures were predicted in the hidden-bottom
dipion decays of $\Upsilon(11020)$ \cite{Chen:2011pu}.

Just illustrated in Eq. (\ref{common}), the similarity existing in
$Y(2175)$, $Y(4260)$ and $\Upsilon(10860)$ stimulates us to apply the ISPE
mechanism to study the hidden-strange dipion decay of $Y(2175)$, which can
result in some novel phenomena in $Y(2175)$. To some extent, carrying out
the search for such phenomena will not only be an important and intriguing
topic, but also provide useful test of the ISPE mechanism.

This work is organized as follows. After the introduction, in Sect.
\ref{sec2} we present the hidden-strange dipion decay of $Y(2175)$ and the
ISPE mechanism. In Sect. \ref{sec3}, the numerical results are given.
Finally, the paper ends with the discussion and conclusion.

\section{The hidden-strange dipion decay of $Y(2175)$ and the ISPE mechanism}\label{sec2}

\begin{figure}[htb]
\centering
\begin{tabular}{cc}
\scalebox{0.8}{\includegraphics{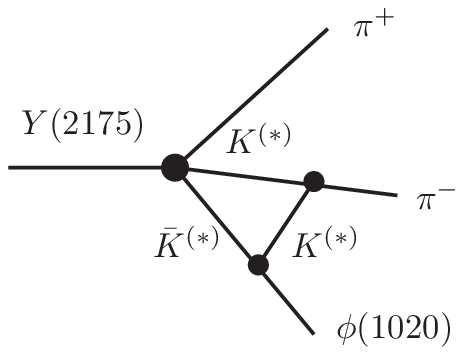}}&
\scalebox{0.8}{\includegraphics{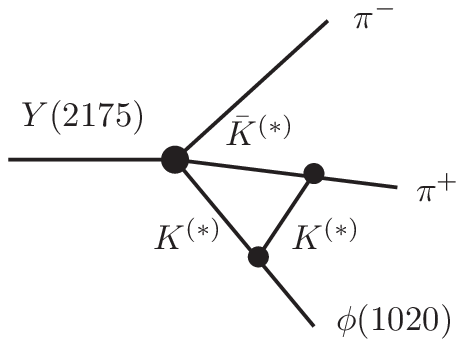}}\\
(a)&(b)
\end{tabular}
\caption{The $Y(2175)\to \phi(1020)\pi^+\pi^-$ decay via the ISPE
mechanism. Here, we only list two schematic diagrams by the initial
$\pi^+$ and $\pi^-$ emissions corresponding to diagrams (a) and (b).}
\label{decay}
\end{figure}

$Y(2175)\to  \phi(1020) \pi^+\pi^-$ occurs via the ISPE mechanism
\cite{Chen:2011pv}, which is depicted in Fig. \ref{decay}. Since the mass
of $Y(2175)$ is just above the thresholds of
$\pi^\pm(K^{(*)}\bar{K}^{(*)})^\mp$, $\pi^\pm$ emitted from $Y(2175)$ is
of continuous energy distribution. Thus, the intermediate $K^{(*)}$ and
$\bar{K}^{(*)}$ can be on-shell or off-shell. What is more important is
that $K^{(*)}$ and $\bar{K}^{(*)}$ with low momenta can easily interact
with each other and then change into $\phi(1020)\pi^\mp$. Since the
minimum of the invariant mass $m_{\phi \pi^\pm}$ is above the $K\bar{K}$ threshold,
in this work we mainly concentrate on the ISPE process
with $K^\ast \bar{K} + \bar{K}^\ast K$ and $K^\ast \bar{K}^\ast$ as the
intermediate states.

Because we use hadron-level description for the hidden-strange dipion
decays of $Y(2175)$, the effective Lagrangian approach
is an appropriate way to describe the decay amplitudes relevant to this process.
The effective interaction Lagrangians
involved in our calculation are given by,
\cite{lin00,Kaymakcalan:1983qq,Oh2000qr}
%
\begin{eqnarray}
  {\cal L}_{K^* K^{(*)}\pi} &=& ig_{K^* K\pi} \bar K^*_\mu \left(\partial^\mu\pi K -
  \pi\;\partial^\mu K\right) + {\rm H.C.}  \nonumber \\
  %
  && -g_{K^* K^*\pi}
  \epsilon^{\mu\nu\rho\sigma}\; \partial_\mu\bar K^*_\nu\;\pi\;
  \partial_\rho K^*_\sigma , \label{eKSKSpi} \\
  {\cal L}_{\phi K^{(*)}K^{(*)}} &=& ig_{\phi KK} \phi^{\mu}\left(\bar K\partial_\mu K -
  {\partial_\mu\bar K} K\right) \label{phiKK} \nonumber \\
  %
  && + g_{\phi K^*K}\epsilon^{\mu\nu\rho\sigma}\;
  \partial_\mu \phi_\nu \left(\bar K\partial_\rho K^*_\sigma + \partial_\rho\bar K^*_\sigma K\right),
  \label{ephiKSK} \nonumber \\
  %
  && +ig_{\phi K^*K^*} \left[\phi^\mu \left(\partial_\mu K^{*\nu}
  \bar K_\nu^* - K^{* \nu} \partial_\mu \bar K_\nu^*\right) \right.
  \nonumber \\
  && +\left(\partial_\mu \phi^\nu K_\nu^* - \phi^\nu \partial_\mu K_\nu^*\right)\bar K^{*\mu}
  \nonumber \\
  && + K^{*\mu} \left.
  \left(\phi^\nu \partial_\mu \bar K_\nu^* - \partial_\mu \phi^\nu \bar K_\nu^*\right)
  \right] , \label{phiKSKS} \\
  {\cal L}_{Y K^{(*)}K^{(*)}\pi} &=&
  %
   -g_{Y K^*K\pi} Y^\mu
  \left(\bar K\pi K^*_\mu+\bar K^*_\mu\pi K\right)
  \label{phiKKpi} \nonumber \\
  %
  && +ig_{Y K^* K^*\pi}\;
  \epsilon^{\mu\nu\rho\sigma}\; Y_\mu \bar K^*_\nu \partial_\rho\pi
  K^*_\sigma,  \nonumber \\
  &&+ ih_{Y K^* K^*\pi}\;
 \epsilon^{\mu\nu\rho\sigma}\left( 2{\partial _\mu }{Y _\nu }{{\bar K}_\rho }^*\pi {K_\sigma }^* \right. \nonumber \\
  &&- \left. {Y _\mu }{\partial _\nu }{{\bar K}_\rho }^*\pi {K_\sigma }^* - {Y _\mu }{{\bar K}_\nu }^*\pi {\partial _\rho }{K_\sigma }^*\right),\label{eYKSKSpi}
\end{eqnarray}
where the terms are derived from the $SU(3)$ invariant effective
Lagrangian, among which those proportional to epsilon tensor are
derived from the Wess-Zumino-Witten Lagrangian that is obtained by gauging
the Wess-Zumino action. The detailed deduction of Eqs.
(\ref{eKSKSpi})-(\ref{eYKSKSpi}) is presented in Appendix.

The coupling constants of the terms derived from the $SU(3)$ invariant Lagrangian have
the following $SU(3)$ limit relations, assuming the vector coupling $g$ for $K^*$ and
$\phi$ is the same but that for $Y$, $g'$ is different from $g$:
\begin{eqnarray*}
  g_{K^* K\pi} &=& \frac{1}{4}~g,\quad
  g_{\phi KK} = \frac{\sqrt{3}}{4}~g \sin\theta, \nonumber \\
  g_{Y K^*K\pi} &=& \frac{\sqrt{3}}{4}~g g', \quad
  g_{\phi K^*K^*} = {-\frac{\sqrt{3}}{4}~{ g}} \sin\theta, \label{Coupling}
\end{eqnarray*}
%
where $\theta$ is a mixing angle between $\omega$ and $\phi$ and is given by $\sin\theta=-0.761$.
The coupling constants proportional to epsilon tensor in the $SU(3)$ limit are expressed by
\cite{Kaymakcalan:1983qq}
\begin{eqnarray*}
  g_{\phi K^*K} &=& \frac{1}{2\sqrt{3}}C_1 \sin\theta, \; 
  \;
  g_{K^* K^*\pi} = \frac{1}{4}C_1,\\ \quad g_{Y K^* K^*\pi} &=& { \frac{1}{2\sqrt{3}}} g^\prime C_1 ,\;
  h_{Y K^* K^*\pi} = { \frac{1}{8\sqrt{3}}} g^\prime C_1 ,
  \label{ECoupling} \\
%
  C_1 &=& \frac{{g^2}N_c}{16\pi^2 F_\pi}, \quad
  C_2 =  \frac{{g}N_c}{6\pi^2 F_\pi^3},
\end{eqnarray*}
with $N_c$ the number of color and {$F_\pi=132$} MeV pion decay constant.

The coupling constant $g$ can be determined by the experimental width of
the process $\phi(1020)\to K^+{K}^-$, i.e., $g=14.9$, where
the total width of $\phi(1020)$ is 4.26 MeV and the branching ratio of
$\phi(1020)\to K^+{K}^-$ is 48.9\% as listed in Particle Data Book
\cite{Nakamura:2010zzi}. On the other hand, we can also determine $g=12.5$
corresponding to the experimental width of $K^*\to K\pi$ ($\Gamma(K^*\to
K\pi)=46.2$ MeV) \cite{Nakamura:2010zzi}, which is consistent with that
obtained by the total width of $\phi(1020)$ mentioned above. In this work, we adopt average value $g=(14.9+12.5)/2=13.7$ when presenting the numerical results. Since we are interested in the
lineshapes of the $d\Gamma(Y(2175)\to \phi(1020)\pi^+\pi^-)/d(m_{\phi
\pi^\pm})$ dependent on $m_{\phi \pi^\pm}$, which are independent on the
value of $g^\prime$.

With these interaction Lagrangians, we write out the decay amplitudes of $Y(2175)\to
(K^{(*)}\bar{K}^{(*)})_{K^{(*)}}^\mp\pi^\pm\to \phi(1020)\pi^+\pi^-$,
where the subscript denotes the corresponding exchanged meson when the
intermediate $(K^{(*)}\bar{K}^{(*)})^\mp$ being transformed into
$\phi(1020)\pi^\mp$. Considering the intermediate $K\bar{K}^*$, $K^*\bar{K}^*$
contributions to $Y(2175)\to \phi(1020)\pi^+\pi^-$, the decay amplitudes
corresponding to Fig. \ref{decay} (a) read ,


\begin{widetext}

\begin{eqnarray}
\mathcal{A}_{K\bar{K}^*}^{{K}^*}(a) &=& \mathcal{I} (i)^3 \int \frac{d^4
q}{(2 \pi)^4} \left[ -g_{YK^\ast K \pi} \epsilon_{Y \mu}\right] \left[
ig_{K^\ast K \pi} (-i p_{4 \rho}- ip_{2 \rho}) \right] \left[ ig_{\phi
K^\ast K^\ast} \epsilon_{\phi}^\nu \big( (-iq_\nu + ip_{2
\nu}) g_{ \lambda \phi} + (i p_{2 \lambda} \right.\nonumber\\
&& \left. + iq_{\lambda}) g_{\nu \phi} + (-ip_{2 \phi}-ip_{5 \phi})g_{\nu
\lambda} \big) \right]  \frac{1}{p_1^2 -m_{K}^2} \frac{-g^{\mu \lambda} +
p_2^\mu p_2^\lambda/m_{K^\ast}^2}{p_2^2-m_{K^\ast}^2} \frac{-g^{\rho \phi}
+ q^\rho q^\phi/m_{K^\ast}^2}{q^2-m_{K^\ast}^2} \mathcal{F}^2(m_{K^\ast}^2,q^2),\label{h1}\\
%
\mathcal{A}_{K^\ast \bar{K}}^{{K}}(a) &=& \mathcal{I}(i)^3 \int \frac{d^4
q}{(2 \pi)^4} \left[-g_{YK^\ast K \pi} \epsilon_{Y \mu}\right] \left[i
g_{K^\ast K \pi} (ip_{4 \rho} -iq_{\rho}) \right] \left[ ig_{\phi KK}
\epsilon_{\phi}^\nu (-iq_{\nu} + ip_{2 \nu}) \right] \nonumber\\
&&\times \frac{-g^{\mu \rho} + p_{1}^{\mu} p_{1}^{\rho} /m_{K^\ast}^2}{
p_1^2- m_{K^\ast}^2} \frac{1}{p_2^2 -m_K^2} \frac{1}{q^2 -m_K^2}
\mathcal{F}^2(m_{K}^2,q^2),\\
%
\mathcal{A}_{K^\ast \bar{K}}^{{K}^*}(a)&=& \mathcal{I}(i)^3 \int \frac{d^4
q}{(2 \pi)^4} \left[-g_{YK^\ast K \pi} \epsilon_{Y \mu}\right] \left[  -g
_{K^\ast K^\ast \pi} \epsilon_{\theta \phi \delta \tau} (-ip_1^\theta)
(i q^\delta)\right] \left[ g_{\phi K^\ast K^\ast} \epsilon_{\rho \nu
\alpha \beta} (ip_5^\nu)
\epsilon_{\phi}^\nu (-iq^\alpha) \right] \nonumber\\
&&\times  \frac{-g^{\mu \phi} +p_1^\mu p_1^\phi/m_{K^\ast}^2}{p_1^2
-m_{K^\ast}^2} \frac{1}{p_2^2 -m_K^2} \frac{-g^{\tau \beta} + q^\tau
q^\beta/m_{K^\ast}^2 }{q^2 -m_{K^\ast}^2} \mathcal{F}^2 (m_{K^\ast}^2q^2),\\
%
\mathcal{A}_{K\bar{K}^*}^{K}(a) &=& \mathcal{I}(i)^3 \int \frac{d^4 q}{(2
\pi)^4} \left[ ig_{Y K^\ast K^\ast \pi} \epsilon_{\mu \rho \alpha
\beta} \epsilon_Y^\mu (i p_3^\alpha) +ih_{Y K^\ast K^\ast \pi}
\epsilon_{\mu \rho \alpha \beta} \epsilon_Y^\mu (-2ip_0^\alpha +
ip_1^\alpha+ ip_2^\alpha)
\right]\left[ ig_{K^\ast K \pi} (ip_{4 \lambda} -iq_{\lambda})\right] \nonumber\\
&&\times \left[g_{\phi K^\ast K} \epsilon_{\theta \nu \delta \tau}
(ip_5^\theta) \epsilon_{\phi}^\nu  (-ip_2^\delta)\right] \frac{-g^{\beta
\lambda}+ p_1^\beta p_1^\lambda/m_{K^\ast}^2 }{p_1^2-m_{K^\ast}^2}
\frac{-g^{\rho \tau} +p_2^\rho p_2^\tau/m_{K^\ast}^2}{p_2^2-m_{K^\ast}^2}
\frac{1}{q^2 -m_K^2} \mathcal{F}^2(m_{K}^2,q^2),\\
\mathcal{A}_{K^*\bar{K}^*}^{{K}^*}(a)&=& \mathcal{I}(i)^3 \int \frac{d^4
q}{(2 \pi)^4}\left[ ig_{Y K^\ast K^\ast \pi} \epsilon_{\mu \rho \alpha
\beta} \epsilon_Y^\mu (i p_3^\alpha) +ih_{Y K^\ast K^\ast \pi}
\epsilon_{\mu \rho \alpha \beta} \epsilon_Y^\mu (-2ip_0^\alpha +
ip_1^\alpha+ ip_2^\alpha)
\right]\nonumber\\
&&\times \left[ -g_{K^\ast K^\ast \pi} \epsilon_{\theta \lambda \delta
\tau} (-ip_1^\theta) (iq^\delta)\right] \left[ ig_{\phi K^\ast K^\ast}
\epsilon^\nu \big( (-iq_\nu +ip_{2\nu})g_{\lambda \phi} + (ip_{5 \lambda}
+ iq_{\lambda})g_{\nu \phi} +(-ip_{2\phi}-ip_{5 \phi})g_{\nu \lambda}
\big) \right]
\nonumber\\
&&\times \frac{-g^{\beta \kappa}+p_1^\beta p_1^\kappa/m_{K^\ast}^2 }{p_1^2
-m_{K^\ast}^2} \frac{-g^{\rho \lambda}+p_2^\rho p_2^\lambda/m_{K^\ast}^2
}{p_2^2 -m_{K^\ast}^2} \frac{-g^{\tau \phi} +q^\tau
q^\phi/m_{K^\ast}^2}{q^2 -m_{K^\ast}^2} \mathcal{F}^2(m_{K^\ast}^2,q^2) ,\label{h6}
\end{eqnarray}
\end{widetext}
where $\mathcal{I}=2$ is due to $SU(2)$ symmetry. The subscript $\mathcal{M}_1 \mathcal{M}_2$ and superscript $\mathcal{M}_3$ in the amplitude
$\mathcal{A}_{\mathcal{M}_1 \mathcal{M}_2}^{\mathcal{M}_3} (a)$ correspond to
the intermediate strange meson pair and the exchanged strange meson, respectively.
By making the
replacements $p_3\rightharpoonup p_4$ and $p_4\rightharpoonup p_3$ in Eqs (\ref{h1})-(\ref{h6}), we can easily obtain the decay amplitudes $\mathcal{A}_{\mathcal{M}_1 \mathcal{M}_2}^{\mathcal{M}_3} (b)$ corresponding to Fig. \ref{decay} (b). Thus, the total decay amplitude of $Y(2175) \to \phi(1020)\pi^+ \pi^-$
is given by
\begin{eqnarray}
\mathcal{A}=\sum_{\mathcal{M}_1 \mathcal{M}_2,\mathcal{M}_3}\Big[\mathcal{A}_{\mathcal{M}_1 \mathcal{M}_2}^{\mathcal{M}_3} (a)+\mathcal{A}_{\mathcal{M}_1 \mathcal{M}_2}^{\mathcal{M}_3} (b)\Big].
\end{eqnarray}
The differential decay width for $Y(2175)\to \phi(1020)\pi^+\pi^-$ reads
\begin{eqnarray}
d\Gamma=\frac{1}{(2\pi)^332m^3_{Y}}\overline{|\mathcal{A}|^2}
d{m^2_{\phi\pi^+}d m^2_{\pi^+\pi^-}},
\end{eqnarray}
where $m_Y$ denotes the mass of $Y(2175)$ and $m_{\pi^+\pi^-}$ is the $\pi^+\pi^-$ invariant mass.
In addition, the overline indicates the average over the polarizations
of the $Y(2175)$ in the initial state and the sum
over the polarization of $\phi(1020)$ in the final state.
In Eqs. (\ref{h1})-(\ref{h6}), we introduce dipole form factor $\mathcal{F}^2(m_E^2,q^2) =((\Lambda^2-m^2_{E})/(\Lambda^2 - q^2))^2$
to reflect the off-shell effects of the exchanged strange meson, which also illustrates the structure effect of interaction vertex between
the intermediated mesons and the exchanged meson. Furthermore, $\mathcal{F}^2(m_E^2,q^2)$ also plays an important role to
make the divergence of loop integral in Eqs. (\ref{h1})-(\ref{h6}) disappear, which is similar to the Pauli-Villas renormalization scheme.
$m_{E}$ in $\mathcal{F}^2(m_E^2,q^2)$ is the mass of the exchanged $K^{(\ast)}$ meson, while
$\Lambda$ is usually parameterized as $\Lambda=m_{E} + \beta
\Lambda_{QCD}$ with Quantum Chromodynamics (QCD) scale $\Lambda_{QCD}= 220$ MeV.

\section{Numerical result}\label{sec3}

With the above preparation, we obtain the distributions of the $\phi(1020) \pi^+$ invariant mass
considering the ISPE mechanism with different strange meson
pairs ($K^\ast\bar{K}+h.c.$ or $K^\ast\bar{K}^\ast$) as the intermediate states. Just because the lineshapes of
the $\phi(1020) \pi^+$ invariant mass distributions are weakly
dependent on the parameter $\beta$ \cite{Chen:2011xk}, in the following we take typical value $\beta=1$ to illustrate
our numerical result.

\begin{figure}[htb]
\scalebox{0.90}{\includegraphics{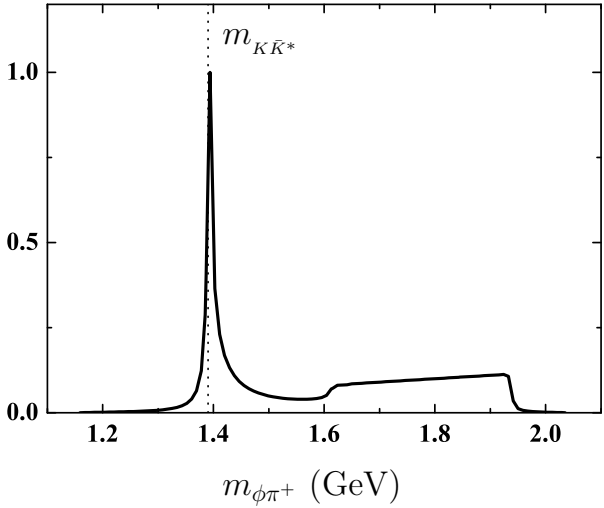}} %
\caption{The lineshape of the $\phi(1020) \pi^+$ invariant mass distribution for
$d\Gamma\left(Y(2175) \to \phi(1020) \pi^+ \pi^-\right)/dm_{\phi\pi^+}$, where $K^{\ast}
\bar{K} +h.c.$ is the intermediate state. The vertical dashed line is the $K\bar{K}^\ast$ threshold. Here, the maximum of the lineshape is
normalized to be $1$. } \label{Fig:KStarK}
\end{figure}

In Fig. \ref{Fig:KStarK}, we present the lineshape of the $\phi(1020) \pi^+$
invariant mass distributions for $Y(2175) \to \phi(1020) \pi^+ \pi^-$, where
the intermediate state is $K^{\ast} \bar{K} +h.c.$. The numerical result indicates that
there should exist a sharp peak near the $K^\ast \bar{K}$ threshold. In addition, we also notice that
the change of lineshape in the 1.62 to 1.94 GeV range becomes smooth, which is due to the contribution from its reflection.

\begin{figure}[htb]
\scalebox{0.90}{\includegraphics{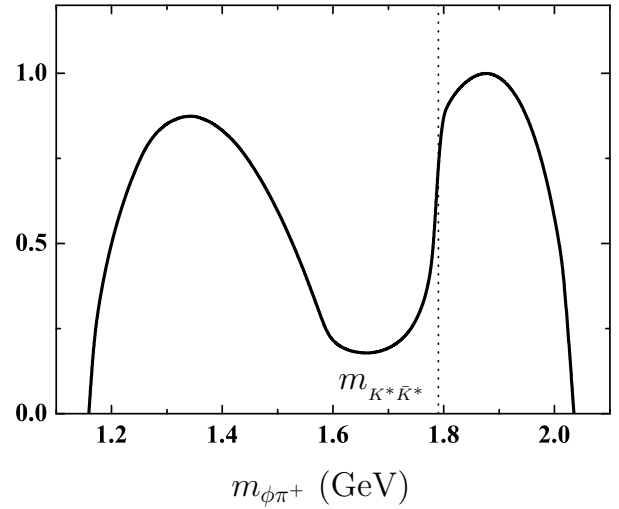}} %
\caption{The lineshape of the $\phi \pi^+$ invariant mass distribution for
$d\Gamma\left(Y(2175) \to \phi \pi^+ \pi^-\right)/dm_{\phi \pi^+}$ resulted from the ISPE mechanism with $K^{\ast}
\bar{K}^{\ast}$ being the intermediate state. Here, the vertical dashed line indicates
the $K^\ast \bar{K}^\ast$ threshold. Additionally, the maximum of the lineshape is also
normalized to be $1$. } \label{Fig:KStarKStar}
\end{figure}

We also present the lineshape of the $\phi(1020) \pi^+$
invariant mass distributions for $Y(2175) \to \phi(1020) \pi^+ \pi^-$ with $K^\ast\bar{K}^\ast$ as intermediate state, which is shown in Fig. \ref{Fig:KStarKStar}. Different from that presented in Fig. \ref{Fig:KStarK}, the lineshape in Fig. \ref{Fig:KStarKStar} indicates that there should exist
two broad structures. One is close to the $K^\ast\bar{K}^\ast$ threshold and another corresponds to its reflection contribution.
To some extent, the difference of the results in Figs. \ref{Fig:KStarK} and \ref{Fig:KStarKStar} reflects different dynamics involved in the
$Y(2175) \to \phi(1020) \pi^+ \pi^-$ channel with the $K^\ast\bar{K}^\ast$ and $K^\ast\bar{K}+h.c.$ intermediate states.

When we applied the ISPE mechanism to the $\Upsilon(5S)\to \Upsilon(1S)\pi^+\pi^-$ process, the result in Ref. \cite{Chen:2011pv} shows that there exists a sharp peak near the $B^*\bar{B}^*$ threshold. However, using the same mechanism to study the $Y(2175)\to \phi(1020)\pi^+\pi^-$ process, one gets a broad structure near the $K^*\bar{K}^*$ just shown in Fig. \ref{Fig:KStarKStar}, which seems to be deviated from the expected result extended from Ref. \cite{Chen:2011pv}. In the following, we find the reasonable explanation to this difference.

\begin{figure}[htb]
\scalebox{0.70}{\includegraphics{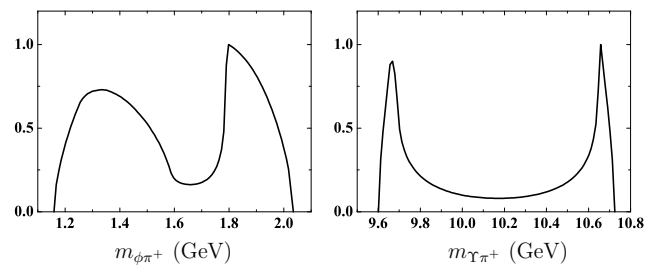}} %
\caption{The lineshapes of the $\phi(1020) \pi^+$ and $\Upsilon(1S) \pi^+$ invariant mass distributions for $Y(2175)\to \phi(1020)\pi^+\pi^-$ and $\Upsilon(5S)\to \Upsilon(1S)\pi^+\pi^-$, respectively. Here, we only consider the intermediate $K^*\bar{K^*}$ and $B^*\bar{B}^*$ contributions to
$Y(2175)\to \phi(1020)\pi^+\pi^-$ and $\Upsilon(5S)\to \Upsilon(1S)\pi^+\pi^-$, respectively. } \label{Fig:Compare}
\end{figure}

If comparing the Lagrangian of the $YK^*K^*\pi$ interaction in Eq. (\ref{eYKSKSpi}) with that of the $\Upsilon(5S)B^*B^*\pi$ coupling adopted in Ref. \cite{Chen:2011pv}, we notice that two extra terms exist in Eq. (\ref{eYKSKSpi}). Here, the Lagrangians presented in this work and in Ref. \cite{Chen:2011pv} are obtained by different approaches. When adopting the same Lorentz structure as that in Ref. \cite{Chen:2011pv} to describe the $YK^*K^*\pi$ interaction, we get the lineshape shown in the left-hand-side diagram of Fig. \ref{Fig:Compare}, where a sharp peak near the $K^*\bar{K}^*$ threshold appears in the $\phi(1020)\pi^+$ invariant mass spectrum, which is different from the result shown in Fig. \ref{Fig:KStarKStar}. This fact further indicates that two extra terms in Eq. (\ref{eYKSKSpi}) play an important role to obtain the result of Fig. \ref{Fig:KStarKStar}.

In addition, we need to emphasize that these two extra Lorentz structures cannot affect the result of $\Upsilon(5S)\to \Upsilon(1S)\pi^+\pi^-$
by the ISPE mechanism. If extending the formulation in Appendices in this paper from SU(3) to SU(5), we can obtain the Lagrangian describing the $\Upsilon(5S)B^\ast B^\ast \pi$ interaction as
\begin{eqnarray}
\mathcal{L}_{\Upsilon B^\ast B^\ast \pi}^{\mathrm{SU(5)}} &=& ig_{\Upsilon B^\ast B^\ast \pi} \varepsilon_{\mu \nu \alpha \beta} \Upsilon^\mu \bar{B}^{\ast \rho} \partial_{\nu} \pi B^{\ast \sigma}  \nonumber\\
&& + ih_{\Upsilon B^\ast B^\ast \pi} \varepsilon_{\mu \nu \alpha \beta}(4  \partial^\nu \Upsilon^\mu \bar{B}^{\ast \rho} \pi B^{\ast \sigma}\nonumber\\&& + \Upsilon^\mu \partial^\nu \bar{B}^{\ast \rho} \pi B^{\ast \sigma} + \Upsilon^\mu \bar{B}^{\ast \rho} \pi \partial^\nu B^{\ast \sigma})\label{new}
\end{eqnarray}
with $g_{\Upsilon B^\ast B^\ast \pi}=6 h_{\Upsilon B^\ast B^\ast \pi}$, where two extra terms also appear compared with the corresponding Lagrangian in Ref. \cite{Chen:2011pv}. If applying this Lagrangian in Eq. (\ref{new}), we obtain
the lineshape of the $\Upsilon(1S) \pi^+$ invariant mass distribution for
$d\Gamma\left(\Upsilon(5S) \to \Upsilon(1S) \pi^+ \pi^-\right)/dm_{\Upsilon(1S) \pi^+}$ just shown in the right-hand-side diagram of Fig. \ref{Fig:Compare}, which is almost the same as that given in Ref. \cite{Chen:2011pv}.

\begin{figure}[htb]
\scalebox{0.90}{\includegraphics{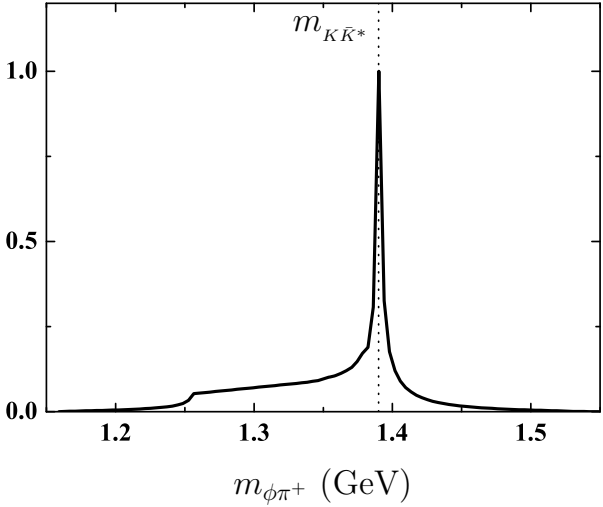}} %
\caption{The lineshape of the $\phi(1020) \pi^+$ invariant mass distribution for
$d\Gamma\left(\phi(1680) \to \phi(1020) \pi^+ \pi^-\right)/dm_{\phi(1020) \pi^+}$, where $K^{\ast}
\bar{K} +h.c.$ is the intermediate state.} \label{1680}
\end{figure}

Apart from studying $Y(2175) \to \phi(1020) \pi^+ \pi^-$ decay, we can easily apply the same formulation to study the $\phi(1680)\to \phi(1020)\pi^+\pi^-$ process, which supplies a suitable platform to study the ISPE mechanism. Here the ISPE mechanism requires that only $K\bar{K}^\ast+h.c.$ be the intermediate state because $m_{\phi(1680)}>m_K+m_{K^\ast}+m_\pi$. The numerical result is given in Fig. \ref{1680}. We also find that a sharp peak structure near the $K\bar{K}^\ast$ threshold appears in the $\phi(1020) \pi^+$ invariant mass distribution. Comparing Fig. \ref{Fig:KStarK} with Fig. \ref{1680}, one notices that the reflection contribution corresponding to the sharp peak is not obvious as shown in Fig. \ref{1680}, which is mainly due to the different phase spaces of
$\phi(1680)\to \phi(1020)\pi^+\pi^-$ and  $Y(2175) \to \phi(1020) \pi^+ \pi^-$ since the Lorentz structures of the decay amplitudes of $\phi(1680)\to \phi(1020)\pi^+\pi^-$ and  $Y(2175) \to \phi(1020) \pi^+ \pi^-$ are exactly the same.

\section{Discussion and conclusion}\label{sec4}

With more and more observations of $XYZ$ states in the past decade, carrying out the study on the properties of these observed hadrons has become an active and intriguing research field. Among these observations,
$Y(2175)$, $Y(4260)$ and $\Upsilon(10860)$ are of similarity just presented in Sec. \ref{sec1}. Due to the recent experimental result of two charged bottomonium-like states $Z_b(10610)$ and $Z_b(10650)$ in $\Upsilon(10860)$ hidden-bottom dipion decays \cite{:2011pd}, the ISPE mechanism, a peculiar decay mechanism existing in the quarkonium dipion hadronic decay, was proposed in Ref. \cite{Chen:2011pv}, which was later applied to study the hidden-charm dipion decays of higher charmonia \cite{Chen:2011xk} and the hidden-bottom dipion decays of $\Upsilon(11020)$ \cite{Chen:2011pu}. Considering the similarity among $Y(2175)$, $Y(4260)$ and $\Upsilon(10860)$, in this work we apply the ISPE mechanism to study the hidden-strange dipion decays of $Y(2175)$ \cite{Aubert:2006bu,:2007yt,Shen:2009zze}, i.e., $Y(2175)\to \phi(1020)\pi^+\pi^-$.

Our calculation shows that a sharp peak structure $Z_{s1}^+$ close to the $K\bar{K}^\ast$ threshold and a broad enhancement $Z_{s2}^+$ around the $K^\ast\bar{K}^\ast$ threshold exist in the $ \phi(1020)\pi^+$ invariant mass spectrum of the $Y(2175)\to \phi(1020)\pi^+\pi^-$ process. We also find a sharp peak structure near the $K\bar{K}^\ast$ threshold when studying $\phi(1680)\to \phi(1020)\pi^+\pi^-$. These predictions of two charged strangeonium-like structures observable in the $Y(2175)\to \phi(1020)\pi^+\pi^-$ and $\phi(1680)\to \phi(1020)\pi^+\pi^-$ processes are accessible at BaBar, Belle II, Supre-B and BESIII. Thus, the experimental search for charged strangeonium-like structures $Z_{s1}^+$ and $Z_{s2}^+$ will be an interesting and important research topic, which will provide further support for the ISPE mechanism.

We can extend the present formulation to study $\omega(1420)\to \omega\pi^+\pi^-$ and $\omega(1650)\to \omega\pi^+\pi^-$ decays if the ISPE mechanism is a universal one existing in quarkonium dipion hadronic transitions. Here, $K\bar{K}^\ast+h.c.$ can be taken as the intermediate state of $\omega(1420)\to \omega\pi^+\pi^-$ and $\omega(1650)\to \omega\pi^+\pi^-$, which should result in the structure appearing in the $\omega\pi^+$ invariant mass spectrum distribution \cite{liu} similar to $Z_{s1}^+$.
It is an effective approach to search for strangeonium-like structure $Z_{s1}^+$ in the $\omega(1420)\to \omega\pi^+\pi^-$ and $\omega(1650)\to \omega\pi^+\pi^-$ processes. Additionally, as shown in Eq. (\ref{common}), there should exist a non-strange partner of $Y(2175)$, which decays into
$\omega\pi^+\pi^-$, where the ISPE mechanism can be applied to study its dipion transition.


\section*{Acknowledgements} This
project is supported by the National Natural Science Foundation of
China under Grant Nos. 11175073, 11005129, 11035006, 11047606,
the Ministry of Education of China (FANEDD under Grant No. 200924,
DPFIHE under Grant No. 20090211120029, NCET, the Fundamental
Research Funds for the Central Universities), the Fok Ying-Tong Education Foundation (No. 131006), and the West
Doctoral Project of Chinese Academy of Sciences.

\section*{Appendix}
\subsection{$SU(3)$ symmetric Lagrangian}

For $SU(3)$ symmetry, we have the following one pseudoscalar and $1^-$
vector multiplet $V$ \cite{lin00}:
%
\begin{eqnarray}
\sqrt{2} P &=& %
\left( {\begin{array}{*{20}{c}}
{\frac{{{\pi ^0}}}{{\sqrt 2 }} + \frac{\eta }{{\sqrt 6 }}}&{{\pi ^ + }}&{{K^ + }}\\
{{\pi ^ - }}&{ - \frac{{{\pi ^0}}}{{\sqrt 2 }} + \frac{\eta }{{\sqrt 6 }}}&{{K^0}}\\
{{K^ - }}&{{{\bar K}^0}}&{ - \sqrt {\frac{2}{3}} \eta}
\end{array}} \right), \label{P3} \\
\sqrt{2}V &=&
\left( {\begin{array}{*{20}{c}}
{\frac{{{\rho ^0}}}{{\sqrt 2 }} + \frac{{\omega_8 }}{{\sqrt 6 }}}&{{\rho ^ + }}&{{K^{* + }}}\\
{{\rho ^ - }}&{ - \frac{{{\rho ^0}}}{{\sqrt 2 }} + \frac{{\omega_8 }}{{\sqrt 6 }}}&{{K^{*0}}}\\
{{K^{* - }}}&{{{\bar K}^{*0}}}&{ - \sqrt {\frac{2}{3}} \omega_8 }
\end{array}} \right) , \label{V3}
%
%
\end{eqnarray}
where
coefficients are determined by normalization and traceless of the matrices $P$ and $V$ octets.
Because a singlet $V^0=\frac{1}{\sqrt{3}}\mathrm{dig}(\phi_1,\phi_1,\phi_1)$ does not have interaction with $P$ and $V$, we exclude this multiplet from hereon
in consideration.
With these multiplets, $P$ and $V$,
we can construct the effective Lagrangian among these particles as \cite{lin00}
\begin{eqnarray}
  D_\mu P &=& \partial_\mu P -\frac{ig}{2} \left[ V_\mu, P\right], \nonumber\\
  {\cal L} &=& {\rm Tr}\left(\left(D_\mu P\right)^\dag D^\mu P\right) -
  \frac{1}{2}{\rm Tr}\left(F_{\mu\nu}^\dag F^{\mu\nu}\right)
  = {\cal L}_0+{\cal L}_{\rm int}, \nonumber \\
  {\cal L}_0 &=& {\rm Tr}\left(\partial_\mu P \partial^\mu P\right) -
  \frac{1}{2}{\rm Tr}\left( \tilde F_{\mu\nu}  \tilde F^{\mu\nu}\right) , \\
  {\cal L}_{\rm int} &=&
  ig {\rm Tr}\left(\partial^\mu P\left[P,  V_\mu\right]\right)
  -\frac{g^2}{4} {\rm Tr}\left(\left[P,  V_\mu\right]^2\right)
  \nonumber\\&&+ ig {\rm Tr}\left(\partial^\mu  V^{\nu}\left[ V_\mu,  V_\nu\right]\right)
  + \frac{g^2}{8} {\rm Tr}\left(\left[ V_\mu,  V_\nu\right]^2\right) , \label{eq_int}
\end{eqnarray}
with $F_{\mu\nu} = \left[D_\mu,D_\nu\right]=
  \partial_\mu  V_\nu - \partial_\nu  V_\mu -
  \frac{ig}{2}\left[ V_\mu,  V_\nu\right]$,
$\tilde F_{\mu\nu} = \partial_\mu  V_\nu - \partial_\nu  V_\mu$,
where  Eq. (\ref{eq_int}) is
obtained because of hermiticity of $P$ and $ V_\mu$.

To derive the interactions among $\phi$ and other particles, we need to consider
the mixing among $\omega$ and $\phi$.
Because of discrepancy between the Gell-Mann-Okubo mass formula and the observed
$\omega$ mass, we normally consider the mixing between $\omega$ and $\phi$ as
\begin{eqnarray}
  \omega_8 = \omega\cos\theta + \phi\sin\theta, \label{mixing}
\end{eqnarray}
where $\sin\theta = -0.761$. See, for instance, pp. 120-121 in Ref. \cite{Chen84}.

The first to the forth terms in Eq. (\ref{eq_int}) correspond to the coupling types
$PPV$, $PPVV$, $VVV$ and $VVVV$ respectively. The concrete Lorentz
structures and coupling constants for $K^\ast K \pi (VP^2)$, $\phi
KK(VP^2)$, $\phi K^\ast K^\ast(V^3)$, $\phi K^\ast K \pi(V^2P^2)$ can be
deduced from the interactions displayed in Eq. (\ref{eq_int})
after inserting Eq.~(\ref{mixing}) into Eq.~(\ref{V3}).

\subsection{$SU(3)$ symmetric Wess-Zumino-Witten action}

The coupling including epsilon tensor structure is obtained by gauging the Wess-Zumino
term, whose general form is given, in terms of differential forms, by \cite{Kaymakcalan:1983qq}
\begin{widetext}
\begin{eqnarray}
  \mathcal{L}_{WZW}\left(U, A_L, A_R\right) &=& iC\int_{M^4} {\rm Tr}\left(A_L~\alpha^3+A_R~\beta^3\right)
  -C\int_{M^4} {\rm Tr}\left[\left(dA_LA_L+A_LdA_L\right)\alpha
  +\left(dA_RA_R+A_RdA_R\right)\beta \right]  \nonumber \\
  &&+ C \int_{M^4} {\rm Tr}\left[dA_L dU A_R U^{-1} - dA_Rd\left(U^{-1}\right) A_L U\right]
  +C \int_{M^4} {\rm Tr}\left(A_RU^{-1}A_LU\beta^2-A_LUA_RU^{-1}\alpha^2\right)
  \nonumber \\
  &&+ \frac{C}{2} \int_{M^4} {\rm Tr}\left[\left(A_L\alpha\right)^2-\left(A_R\beta\right)^2\right]
  +iC \int_{M^4} {\rm Tr}\left(A_L{}^3\alpha+A_R{}^3\beta\right)
  \nonumber \\
  &&+ iC \int_{M^4} {\rm Tr}\left[\left(dA_RA_R+A_RdA_R\right)U^{-1}A_LU
  -  \left(dA_LA_L+A_LdA_L\right)UA_RU^{-1}\right]
  \nonumber \\
  &&+ iC \int_{M^4} {\rm Tr}\left(A_L U A_R U^{-1}A_L\alpha+A_R U^{-1} A_L UA_R\beta\right)
  \nonumber \\
  &&+ C \int_{M^4} {\rm Tr}\left[A_R{}^3 U^{-1} A_L U-A_L{}^3 U A_R U^{-1}
  +\frac{1}{2}\left(UA_RU^{-1}A_L\right)^2\right]
  - Cr \int_{M^4} {\rm Tr}\left(F_LUF_RU^{-1}\right),
\end{eqnarray}
\end{widetext}
where
\begin{eqnarray}
  C &=& -\left(5iN_c\right)/\left(240{\pi^2}\right), \quad V =  gV_\mu dx^\mu,
 \nonumber\\ U&=&\exp\left(2iP/\left(F_\pi\right)\right),\quad
  \alpha = \left(\partial_\mu U\right)U^{-1}dx^\mu
  \equiv (dU)U^{-1}, \nonumber\\ \beta &=& U^{-1}dU=U^{-1}\alpha U,
  \nonumber \\
  A_L &=& \frac{1}{2}(V+A),\quad A_R = \frac{1}{2}(V-A), \nonumber
\end{eqnarray}
from which one can construct higher forms.
Here $P$, $V$ and $A$ are pseudoscalar, vector and axial vector fields,
respectively.

This action with $r=0$ gives the following interaction action with the epsilon tensor structure as \cite{Oh2000qr}:
\begin{eqnarray}
  && \int d^4x{\cal L}_{WZW} = -\frac{g^2N_c}{16\pi^2F_\pi} \int_{M^4} {\rm Tr}
  \left(\left(dV\right)^2P\right)  \nonumber \\
  &&\quad - \frac{igN_c}{6\pi^2{F_\pi^3}} \int_{M^4} {\rm Tr}\left[V(dP)^3\right]
  +\frac{ig^3N_c}{32\pi^2F_\pi} \int_{M^4} {\rm Tr}\left(V^3dP\right)
  \nonumber \\
  &&\quad + \frac{ig^3N_c}{32\pi^2F_\pi} \int_{M^4} {\rm Tr}\left(VdVVP\right), \label{LintE}
\end{eqnarray}
where we have dropped the $A$ field in $A_{L/R}$, hence $A_L=A_R=V$ and all
terms are four-forms because $V$ is one-form and $P$ is zero-form,
i.e.,
\begin{eqnarray}
  \int_{M^4} {\rm Tr}\left(\left(dV\right)^2P\right) &=&
  \epsilon^{\mu\nu\rho\sigma} \int d^4x
  {\rm Tr}\left(\partial_\mu V_\nu \partial_\rho V_\sigma P\right),\\
  \int_{M^4} {\rm Tr}\left[V(dP)^3\right] &=&
  \epsilon^{\mu\nu\rho\sigma} \int d^4x
  {\rm Tr}\left(V_\mu\partial_\nu P\partial_\rho P\partial_\sigma P\right), \\
  \int_{M^4} {\rm Tr}\left(V^3dP\right) &=&
  \epsilon^{\mu\nu\rho\sigma} \int d^4x
  {\rm Tr}\left(V_\mu V_\nu V_\rho\partial_\sigma P\right) , \\
  \int_{M^4} {\rm Tr}\left(VdVVP\right) &=&
  \epsilon^{\mu\nu\rho\sigma} \int d^4x
  {\rm Tr}\left(V_\mu \partial_\nu V_\rho V_\sigma P\right).
\end{eqnarray}
where the left hand sides are written in terms of forms but the right hand sides are in terms of
matrices given by Eqs.~(\ref{P3}-\ref{V3}).

From the above forms of interaction, we easily obtain the couplings $\phi
K^*K$, $\phi K K\pi$, $\phi K^*K^*\pi$ and $K^*K^*\pi$,  which are
classified into $V^2P$, $VP^3$, $V^3P$, and $V^2P$, respectively.

In the above expressions for interaction, the four-point vertices $\phi
K^{(*)}K^{(*)}\pi$ should read $Y K^{(*)}K^{(*)}\pi$ to be used in our
model given by Eq.~(\ref{eYKSKSpi}).
More care has to be taken that the coupling for $Y$, $g'$ is different
from $g$ for $\phi$ and $K^*$ when calculating the interaction.

\end{document}